\documentstyle[aps,prd,epsfig]{revtex}
\begin{document}
\title{FINITE TEMPERATURE INDUCED FERMION NUMBER\footnote{Based on talks
given at Workshop on Lattice Hadron Physics (Cairns, July 2001),
Workshop on Light-Cone: Particles and Strings (Trento, September 2001),
and Workshop on Quantum Field Theory in External Conditions (Leipzig,
September 2001).}}
\author{Gerald V. Dunne}
\address{Department of Physics, University of Connecticut, Storrs, CT
06269-3046, USA}
%\date{\today}
\maketitle
%\vskip .5cm

\vspace*{0.21truein}
\begin{abstract}The induced fractional fermion number at zero temperature
is topological (in the sense that it is only sensitive to global
asymptotic properties of the background field), and is a sharp observable
(in the sense that it has vanishing rms fluctuations). In contrast, at
finite temperature, it is shown to be generically nontopological, and 
not a sharp observable.
\end{abstract}

\section{Introduction}

The phenomenon of induced fermion number arises due to the interaction of
fermions with nontrivial topological backgrounds (e.g., solitons, vortices,
monopoles, skyrmions), and has many applications ranging from polymer physics
to particle physics \cite{jr,gw,bag,niemi,polymer}. At zero
temperature, the induced fermion  number is a topological quantity, and
is related to the spectral asymmetry of the relevant Dirac operator,
which counts the difference between the number of positive and negative
energy states in the fermion spectrum. Mathematical results, such as index
theorems and Levinson's theorem, imply that the zero temperature
induced fermion number is determined by the asymptotic topological properties
of the background fields \cite{boy,mike,alexios,jaffe}. This topological
character of the induced fermion number is a key feature of its
application in certain model field theories. At finite  temperature, the
situation is very different \cite{ad,dr}. In this talk I explain why at
finite T the induced fermion number is generically nontopological, and is
not a sharp observable. 

This work has also been partly motivated by recent results that certain
anomalies in $T=0$ field theory become much more
subtle at nonzero $T$. For example \cite{pisarski}, in even dim.
spacetime, finite T anomalous $\pi^0$ decay amplitudes are T dependent
even though the chiral anomaly (a topological object that is related to
the anomalous $\pi^0$ decay amplitude at $T=0$) is independent of $T$. 
And, in odd dim. spacetime, the topological Chern-Simons term is the
only parity-violating term induced in the $T=0$ effective
action, but at finite $T$ there are infinitely many
parity-violating terms \cite{dll,deser1,schaposnik1}. 

\section{Induced fermion number}

The induced fermion number is an expectation value of the fermion number
operator, $N=\int dx\,\frac{1}{2}[\Psi^\dagger, \Psi]$. For a
given classical static background field configuration, the
fermion field operator $\Psi$ can be expanded in a complete set of
eigenstates of the Dirac Hamiltonian $H$. The $T=0$
fermion number, $\langle N\rangle_0\equiv\langle 0|N|0\rangle$, is
related to the spectral asymmetry of the Dirac Hamiltonian:
\begin{eqnarray}
\langle N\rangle_0=-\frac{1}{2}\int_{-\infty}^\infty dE\, \sigma(E)\,
{\rm sign}(E)
=-\frac{1}{2}\,({\rm spectral~~asymmetry}).
\label{asymmetry}
\end{eqnarray}
Here $\sigma(E)$ is the spectral function of $H$:
\begin{eqnarray}
\sigma(E)=\frac{1}{\pi}\, {\cal I}m\, {\rm
Tr}\left(\frac{1}{H-E-i\epsilon}\right).
\label{spectral}
\end{eqnarray}
So, $\langle N\rangle_0$  essentially counts the number of positive
energy states minus the number of negative energy states. 

At nonzero temperature $T$, the induced fermion number is a {\it thermal}
expectation value
\begin{eqnarray}
\langle N\rangle_{T}={{\rm Tr}\left(e^{-\beta H}\, N
\right) \over {\rm Tr}\left(e^{-\beta H}\right)}
=-\frac{1}{2}\int_{-\infty}^\infty dE\, \sigma(E)\, {\rm
tanh} \left(\frac{\beta E}{2}\right)
\label{nt}
\end{eqnarray}
where $\beta=\frac{1}{T}$. This finite T expression (\ref{nt})
reduces smoothly to the zero T expression (\ref{asymmetry}) as
$\beta\to\infty$; in fact, finite temperature provides a physically
meaningful and mathematically elegant regularization of the spectral
asymmetry.

Note that the spectrum of the fermions in the static background has
nothing to do with the temperature. All information about the fermion
spectrum is encoded in the spectral function. Thus, if one knows
$\sigma(E)$, one immediately has integral representations for both 
$\langle N\rangle_{0}$ and $\langle N\rangle_{T}$. In fact, one only
needs to know the odd part of $\sigma(E)$, as is clear from
(\ref{asymmetry}) and (\ref{nt}). It is convenient to use the explicit
form (\ref{spectral}) of the spectral function to write (\ref{nt}) as a
contour integral involving the resolvent of the Dirac Hamiltonian
\begin{eqnarray}
\langle N\rangle_{T}=-\frac{1}{2} \int_{\cal C}\frac{dz}{2\pi i}\, {\rm
Tr}\left(
\frac{1}{H-z}\right)\, {\rm tanh}\left(\frac{\beta z}{2}\right)
\label{contour}
\end{eqnarray}
where ${\cal C}$ is the contour $(-\infty+i\epsilon,+\infty+i\epsilon)$
and $(+\infty-i\epsilon,-\infty-i\epsilon)$ in the complex energy plane.
Thus the resolvent is the key to the computation.

\section{Two 1+1 dimensional examples}

Before going to the general case, it is instructive to consider two
important cases in 1+1 dim field theory. These two examples are
very similar at T=0, but turn out to be very different for $T>0$. Consider
an abelian model of fermions interacting via scalar
and pseudoscalar couplings to two (classical, static) bosonic fields
$\phi_1$ and $\phi_2$: 
\begin{eqnarray}
{\cal L}=i\,\bar{\psi}\partial \hskip -5pt/\,\psi -
\bar{\psi}\left(\phi_1+i\, \gamma_5\,\phi_2\right)\psi
\label{lag}
\end{eqnarray}
We distinguish between two different cases:

(a) kink case \cite{jr}: $\phi_1=m$ is constant, but
$\phi_2(x)$ has a kink-like shape (for example $\phi_2(x)={\rm
tanh}(x)$), with $\phi_2(\pm\infty)=\pm\hat{\phi_2}$.

(b) sigma-model case \cite{gw}: $\phi_1(x)$ and $\phi_2(x)$ are
constrained to the ``chiral circle'': $\phi_1^2+\phi_2^2=m^2.$

The kink case (a) is relevant to the polymer physics applications, while
the sigma-model case (b) is a low-dimensional analogue of sigma models
used in 3+1 dim. particle physics models \cite{niemi}.

In each case, we define the angular field
\begin{eqnarray}
\theta(x)\equiv {\rm arctan}\,\left[\phi_2(x)/\phi_1(x)\right]
\label{theta}
\end{eqnarray}
In the sigma model case (b), the angular field
$\theta(x)$ has the interpretation of a local chiral angle, and is chosen
to have a kink-like shape, with asymptotic values
$\theta(\pm\infty)\equiv\pm\hat{\theta}$. Note that in the kink case
(a), the angular field $\theta(x)={\rm arctan}[\phi_2(x)/m]$
also has a kink-like shape. 

For {\it both} the kink case and the sigma-model case, at zero T, the
induced fermion number is 
\begin{eqnarray}
\langle N\rangle_{0}=\frac{1}{2\pi}\int dx\,\theta^\prime(x)=
\frac{\hat{\theta}}{\pi}
\label{nt1}
\end{eqnarray}
which is {\it topological} in the sense that it only depends on the
asymptotic values of $\theta(x)$, not on its detailed shape. In the kink
case, this calculation is easy because the even part of the
corresponding Dirac resolvent is known exactly \cite{callias}
\begin{eqnarray}
\left[{\rm Tr}\left(\frac{1}{H-z}\right)\right]_{\rm e} 
={- m \hat{\phi}_2 \over (m^2-z^2)\,\sqrt{m^2+\hat{\phi}_2^2-z^2}}
\label{callias}
\end{eqnarray}
But in the sigma-model case the resolvent is not known exactly, and one
must resort to an approximation such as the derivative
expansion. Nevertheless, both cases lead to the same result
(\ref{nt1}).

At finite T, the kink case is also easy because we know the relevant part
of the resolvent exactly \cite{ns}. Inserting (\ref{callias}) into the
contour integral expression (\ref{contour}) one finds 
\begin{eqnarray}
\langle N\rangle_{T}^{\rm kink}=\sum_{n=0}^\infty {
\frac{2}{\pi}\,(\frac{m\beta}{\pi})^2  {\rm sin}\hat{\theta}\over
((2n+1)^2 +(\frac{m\beta}{\pi})^2)
\sqrt{(2n+1)^2 {\rm cos}^2\hat{\theta}+(\frac{m\beta}{\pi})^2}}
\label{kt}
\end{eqnarray}
Note that $\langle N\rangle_{T}^{\rm kink}$ is temperature dependent, but
it is still topological in the sense that it only depends on the
background field through its asymptotic value $\hat{\theta}={\rm
arctan}[\hat{\phi_2}/m]$. Also, even though the expression (\ref{kt})
looks complicated, it smoothly reduces to (\ref{nt1}) as $T\to 0$. It
is also interesting that the angular nature of $\hat{\theta}$ is more
obvious at finite T than at zero T.

At finite T, the sigma-model case is much more complicated, because there
is no simple closed formula for the relevant part of the resolvent. In
\cite{ad} the derivative expansion was used, and it was shown that for
$T\ll m$ the derivative expansion can be resummed to all orders,
leading to the following simple expression
\begin{eqnarray}
\langle N\rangle_T^{\rm sigma}=\langle N\rangle_0^{\rm sigma}
-\sqrt{\frac{2m}{\beta\pi}}\int dx e^{-m\beta}\, 
\sinh\left(\frac{\beta\theta^\prime}{2}\right)+\dots
\label{st}
\end{eqnarray}
The correction term is temperature dependent and nontopological, as it
is sensitive not just to the asymptotic values of $\theta(x)$, but also to
the details of the shape of $\theta(x)$. This nontopological term has a
simple physical interpretation as follows \cite{ad}. For the sigma model
case we can write the interaction term in (\ref{lag}) as $m
\bar{\psi}\left(
\cos\theta+i\gamma_5 \sin\theta\right)\psi=m \bar{\psi}\, e^{i\gamma_5
\theta} \psi$. Then make the local chiral rotation $
\psi=e^{-i\gamma_5\theta/2}\,\tilde{\psi}$ to obtain 
\begin{eqnarray}
{\cal L}=i\bar{\tilde{\psi}}\partial\hskip -5pt / \tilde{\psi}-
m \bar{\tilde{\psi}} \tilde{\psi}
+\bar{\tilde{\psi}}\, \gamma^0\, \frac{\theta^\prime}{2}\,\tilde{\psi}
\label{rotlag}
\end{eqnarray}
which describes a fermion of mass $m$ in an inhomogeneous electric field  
$E(x)=-\frac{1}{2}\theta^{\prime\prime}(x)$. In this language the zero T
induced fermion number (\ref{nt1}) can be understood as the vacuum
polarization, due to the inhomogeneous electric field background, of
virtual vacuum $e^+e^-$ dipoles. This effect is independent of the
temperature as the virtual vacuum dipoles do not live long enough to
thermalize. However, at finite T there is another effect of the
background electric field. The real particles and antiparticles respond
to this electric field in a way that can be described by linear response
theory. In the derivative expansion limit, where
$\theta^\prime\ll m$, the leading effect of the electric field background
can be represented as a local chemical potential
$\mu(x)=-\frac{1}{2}\theta^\prime(x)$.
With local Fermi distributions
\begin{eqnarray}
f_\pm(x,k)={1\over e^{\beta(\sqrt{k^2+m^2}\,\mp\mu(x))}+1}
\label{fermi}
\end{eqnarray}
the linear response charge density is
$\rho(x)=\int \frac{dk}{2\pi}\, f(x,k)$,
where $f=f_+-f_-$. For $T\ll m$, 
\begin{eqnarray}
\rho(x)\sim -\sqrt{\frac{2mT}{\pi}}\, e^{-m/T}\, 
\sinh\left(\frac{\theta^\prime}{2T}\right)
\label{lr}
\end{eqnarray}
in perfect agreement with the non-topological term in (\ref{st}).

So, while at zero T, the kink and sigma-model cases each have induced
fermion number given by the same expression (\ref{nt1}), at nonzero T they
have very different induced fermion number, given by (\ref{kt}) and
(\ref{st}) respectively. In each case $\langle N\rangle_T$ is temperature
dependent, but in the kink case it is topological, while in the
sigma-model case it is nontopological. The physical reason for this
difference is explained in the next section.

\section{Topological versus Nontopological}

The straightforward mathematical reason why $\langle N\rangle_T$ is
topological for the kink case is that the odd part of the spectral
function is itself topological, as shown by the Callias result
(\ref{callias}). But a much more physical understanding can be
obtained by considering the symmetries of the fermionic spectrum
\cite{dr}. And this perspective generalizes immediately to higher
dimensional models.

Begin with the simple trigonometric identity, $\tanh(\frac{\beta
E}{2})=1-2 n(E)$, where
$n(E)=1/(e^{\beta E}+1)$ is the Fermi-Dirac distribution function. Then 
the finite T fermion number (\ref{nt}) separates as 
\begin{eqnarray}
\langle N\rangle=\langle N\rangle_0 +\int_{-\infty}^\infty
dE\, \sigma(E)\, {\rm sgn}(E)\, n( |E|)
\label{separation}
\end{eqnarray}
The first term is just the zero temperature contribution, which can be
expressed in terms of the spectral asymmetry, and is known to be
topological. But the second term is highly sensitive to the details of
the spectrum, because of the presence of the Fermi weighting factor 
$n(|E|)$. Thus, the only way this term can be topological is if the odd
part of $\sigma(E)$ is itself topological. In terms of the spectrum, this
can only happen if the spectrum is symmetric under $E\to -E$. Remarkably,
this does sometimes happen, but only for extremely special backgrounds.
One example of such a special background is the one-dim kink
background discussed in the previous section. Indeed, for any kink
background (with $\hat{\theta}$ positive), the spectrum of the Dirac
Hamiltonian always has the form shown in Fig. 1. 
%\begin{figure}[htbp] %ORIGINAL SIZE: width=1.4TRUEIN; height=1.5TRUEIN
%\vspace*{13pt}
%\centerline{\psfig{file=f8.eps}} %100 percent
%\vspace*{13pt}
%\fcaption{The general structure of the Dirac spectrum for a kink
%background.}
%\end{figure}
\begin{figure}[h]
\centerline{\includegraphics[scale=0.8]{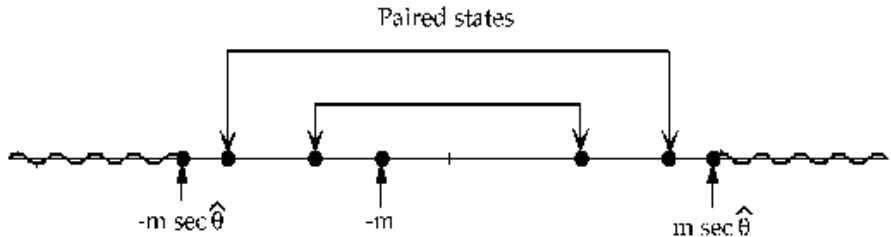}}
\caption{The general structure of the Dirac spectrum for a kink
background.}
\label{f8}
\end{figure}
There are continuum states for $|E|>\sqrt{m^2+\hat{\phi}_2^2}=m\, {\rm
sec}\,\hat{\theta}$.  In addition, there is always a bound state at
$E=-{\rm sign}(\hat{\theta})\,m$. There may or may not be additional
bound states with $m<|E|<m\,{\rm sec}\,\hat{\theta}$. But if these
additional bound states are present, they necessarily occur in $\pm E$
pairs, because of the quantum mechanical SUSY of the Dirac Hamiltonian in
the kink case \cite{niemi}. Thus, the contributions of these paired
bound states to the integral in (\ref{separation}) cancel in pairs, while
the unpaired bound state at $E=-{\rm sign}(\hat{\theta})\,m$ leads to a
contribution $-{\rm sign}(\hat{\theta})\,n( m)$. The quantum mechanical
SUSY of the Dirac Hamiltonian is likewise the key to the exact Callias
index theorem result \cite{callias} for the odd part of the spectral
function [which is equivalent to knowing the even part of the resolvent in
(\ref{callias})]. This means that the remaining integral over the
continuum states can be expressed as an integral of the odd part of the
spectral function, weighted by the Fermi factor $n(E)$, over the positive
continuum beginning at the threshold energy $m{\rm sec}\hat{\theta}$. 
Putting all this together, we obtain
\begin{eqnarray}
\langle N\rangle_T^{\rm kink}=\langle N\rangle_0^{\rm kink}-
n(m) +
\int_{m\,{\rm sec}\,\hat{\theta}}^\infty \frac{dE}{\pi}\, \frac{2 m^2
\tan\hat{\theta}\, n(
E)}{(E^2-m^2)\sqrt{E^2-m^2{\rm sec}^2\hat{\theta}}}
\label{sw}
\end{eqnarray}
One can check that this is, in fact, the Sommerfeld-Watson transform of
the summation expression found before in (\ref{kt}).
%\begin{figure}[htbp] %ORIGINAL SIZE: width=1.4TRUEIN; height=1.5TRUEIN
%\vspace*{13pt}
%\centerline{\psfig{file=f9.ps}} %100 percent
%\vspace*{13pt}
%\fcaption{The structure of the Dirac spectrum for a sigma model
%background.}
%\end{figure}
\begin{figure}[b]
\centerline{\includegraphics[scale=0.8]{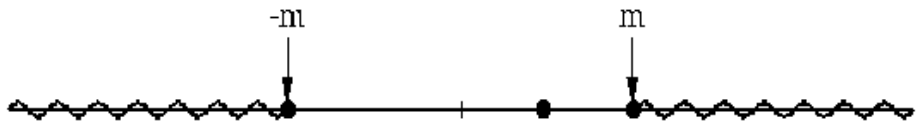}}
\caption{The structure of the Dirac spectrum for a sigma model
background.}
\label{f9}
\end{figure}

On the other hand, for the sigma-model case the fermion spectrum has no
obvious symmetry. In this case, the Dirac spectrum has the form shown in
Fig. 2. The only general thing we can say is that the continuum thresholds
are at $E=\pm m$. There may or may not be one or more bound states for
$|E|<m$. The existence of (and the precise location of) bound states is
highly sensitive to the actual {\it shape} of the $\theta(x)$. As there
is no special symmetry in the spectrum, there is nothing general that can
be said about the odd part of the spectral function. Nevertheless, the
first term in (\ref{separation}), which is $\langle N\rangle_0$, yields a
topological result since the integral can be written, using Levinson's
theorem, in terms of the phase shift at infinity, which is topological
\cite{boy,mike,jaffe}. This application of Levinson's theorem
doesn't work for the T dependent second term in (\ref{separation}) because
of the Fermi factor. This is another way to understand why the finite
temperature induced fermion number is generically nontopological. 

These arguments based on the symmetry of the fermion spectrum (in the
given background) clearly generalize to any dimension of spacetime. So,
we conclude that the generic situation is that the finite T induced
fermion number is non-topological; but, if the background is such that
the fermion spectrum is highly symmetric, then the induced fermion number
may be topological even though it is temperature dependent.

\section{Thermal fluctuations of fermion number}

Fluctuations in the induced fermion number can be analyzed in an
analogous manner \cite{dr}: 
\begin{eqnarray}
(\Delta N)^2\equiv \langle N^2\rangle-\langle
N\rangle^2=\frac{1}{4}\int_{-\infty}^\infty dE\, \sigma(E)\, {\rm sech}^2 
\left(\frac{\beta E}{2}\right)
\label{fluc1}
\end{eqnarray}
The trigonometric identity, $\frac{1}{4}{\rm sech}^2(\frac{\beta
E}{2})=n(E)(1-n(E))$, re-expresses this as
\begin{eqnarray}
(\Delta N)^2=\int_{-\infty}^\infty dE\,
\sigma(E) \, n( E)\left(1-n(E)\right)
\label{fluc}
\end{eqnarray}
which is simply a sum over the fermion spectrum of $n(1-n)$, as it should
be for the thermal occupation of independent single-particle states
\cite{pathria}. From (\ref{fluc}) it is clear that the fluctuations are
always non-topological, as they are sensitive to the details of the
fermion spectrum. Another way to see this is to note that the expressions
(\ref{fluc1},\ref{fluc}) require knowledge of the even part of the
spectral function, which is not accessible to an index theorem. So the
fluctuations must typically be calculated approximately; however, for a
special class of kink backgrounds, $\phi_2(x)=\hat{\phi}
\tanh(\hat{\phi}x/j)$, where $j$ is an integer, it is possible to compute
$(\Delta N)^2$ exactly, and one sees the explicit
(nontopological) dependence on the scale factor $j$ \cite{dr}.

These expressions also make it clear why the fluctuations vanish at zero
T, because the ${\rm sech}^2(\beta E/2)$ factors vanish exponentially fast
as $T\to 0$. Thus, the induced fermion number is a sharp
observable at $T=0$, in agreement with previous results \cite{kivelson},
but at nonzero T it is not a sharp observable, simply because the thermal
expectation value mixes states other just the ground state. Another
example of non-sharp fractional fermion number (but not in the context of
temperature dependence) has been discussed recently in the context of
liquid helium bubbles \cite{helium}.

\section{Conclusions}
To conclude, the finite temperature induced fermion number
$\langle N\rangle_{T}$ is generically nontopological, and is not a
sharp observable. The induced fermion number naturally splits into a
topological temperature-independent piece that represents the effect of
vacuum polarization on the Dirac sea, and a nontopological
temperature-dependent piece that represents the thermal
population of the available states in the fermion spectrum, weighted with the
appropriate Fermi-Dirac factors. As the temperature approaches zero, the
nontopological terms vanish exponentially fast, and we regain smoothly the
familiar topological results at zero temperature. But at nonzero temperature,
the nontopological contribution is more sensitive to the details of the
spectrum, and so is generically nontopological. This argument applies in
any dimension.

Consider, for example,
for a chiral $SU(2)$ sigma model with a Skyrme background in 3+1 dim., 
\begin{eqnarray}
{\cal L}_{\rm int}&=&m\bar{\psi}\left(\pi_0+i\gamma_5 \vec{\pi}\cdot\vec{\tau}
\right)\psi \nonumber\\ 
&=& m\bar{\psi}\left(\frac{1}{2}(g+g^\dagger)+
\frac{1}{2}(g-g^\dagger)\gamma_5\right)\psi
\nonumber
\end{eqnarray}
where $\pi_0^2+\vec{\pi}^2=1$ and $\pi_0+i\vec{\pi}\cdot \vec{\tau}=g$,
which is the analogue of the 1+1 dim. sigma-model case analyzed in this
paper. From numerical work \cite{ripka}, the Dirac
energy spectrum of the fermions is not symmetric, and the energy of a
possible bound state (and indeed the number of bound states) is highly
sensitive to  the details of the shape of the radial hedgehog field.
This shows that in this case the finite temperature induced fermion number
is nontopological. Thus, for a (large) Skyrme background in 3+1 dim., the
finite temperature fermion number is not simply the (topological) winding
number of the Skyrme field, but a much more complicated nontopological
object.

\bigskip
\noindent{\bf Acknowledgement:}
This talk is based on work done in collaboration with Ian Aitchison and
Kumar Rao. I also thank the US DOE for support.

%\nonumsection{References}

\end{document}